\documentclass[journal=jctcce,manuscript=article]{achemso}

\usepackage[version=3]{mhchem} 

\author{Michele Invernizzi}
\affiliation{Department of Physics, ETH Zurich c/o USI Campus, Lugano, Switzerland}
\altaffiliation{Facolt\`{a} di Informatica, Instituto di Scienze Computationali, and National Center for Computational Design and Discovery of Novel Materials MARVEL, Universit\`{a} della Svizzera italiana (USI), Via Giuseppe Buffi 13, CH-6900 Lugano, Switzerland}
\email{michele.invernizzi@phys.chem.ethz.ch}

\author{Michele Parrinello}
\affiliation{Department of Chemistry and Applied Biosciences, ETH Zurich c/o USI Campus, Lugano, Switzerland}
\altaffiliation{Facolt\`{a} di Informatica, Instituto di Scienze Computationali, and National Center for Computational Design and Discovery of Novel Materials MARVEL, Universit\`{a} della Svizzera italiana (USI), Via Giuseppe Buffi 13, CH-6900 Lugano, Switzerland}

\title{Making the best of a bad situation: \\
a multiscale approach to free energy calculation}

\begin{document}

\begin{tocentry}
  \includegraphics[width=4.9cm]{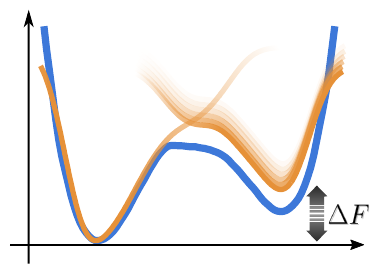}
\end{tocentry}

\begin{abstract}
   Many enhanced sampling techniques rely on the identification of a number of collective variables that describe all the slow modes of the system.
   By constructing a bias potential in this reduced space one is then able to sample efficiently and reconstruct the free energy landscape.
   In methods like metadynamics, the quality of these collective variables plays a key role in convergence efficiency.
   Unfortunately in many systems of interest it is not possible to identify an optimal collective variable, and one must deal with the non-ideal situation of a system in which some slow modes are not accelerated.
   
   We propose a two-step approach in which, by taking into account the residual multiscale nature of the problem, one is able to significantly speed up convergence.
   To do so, we combine an exploratory metadynamics run with an optimization of the free energy difference between metastable states, based on the recently proposed variationally enhanced sampling method.
   This new method is well parallelizable and is especially suited for complex systems, because of its simplicity and clear underlying physical picture.
\end{abstract}

\section{Introduction}
Many systems are characterized by metastable states separated by kinetic bottlenecks.
Examples of this class of phenomena are chemical reactions, first order phase transitions and protein folding.
Calculating the difference in free energy between these states is of great importance and various methods have been developed to reach this goal\cite{Chipot2007,Abrams2014}.
In this paper we focus on the case in which only two states are relevant.
The generalization to a multistate scenario is not too difficult and will be discussed in a future publication.

In the two-state case one can distinguish two time scales, a shorter one in which the system undergoes fluctuations while remaining in one of the minima, and a longer one in which the system moves from one minimum to another.
This is the so-called rare event scenario, in which the separation between the two time scales can be so large as to be inaccessible to direct molecular dynamics simulation, preventing a direct calculation of the free energy surface (FES).

In order to address this issue a widely used class of free energy methods, including, among others, umbrella sampling\cite{Torrie1977}, metadynamics\cite{MetaD} (MetaD) and variationally enhanced sampling\cite{Valsson2014,Valsson2018} (VES), aim at closing this timescale gap by speeding up the sampling of the slow modes of the system.
To achieve this result one first identifies a set of order parameters or collective variables (CVs), that are functions of the atomic coordinates $\mathbf{s}=\mathbf{s}(\mathbf{R})$.
The CVs describe the slow modes whose sampling is accelerated by means of an applied bias potential $V(\mathbf{s})$.
An optimal CV is such that, if employed for an enhanced simulation run, it closes the timescale gap, so that intra-state and inter-states exploration take place on the same timescale.
A more common scenario is that the CV does not encode some slow modes and although much reduced, the gap remains also in the enhanced sampling run.
We shall refer to such CV as suboptimal.
The limiting case would be that of a CV so poorly chosen that the gap falls outside the computational reach.
In this case the CV would be not just suboptimal, but also a bad CV.
Given the crucial role of CVs, much effort has been devoted to their design and their systematic improvement\cite{Branduardi2007,Pietrucci2011,Tiwary2016,McCarty2017,Mendels2018,Pfaendtner2015}.

In metadynamics and variationally enhanced sampling the effect of suboptimal CVs manifests itself in a hysteretic behaviour, easily detectable in the CV evolution.
In these cases MetaD and VES efficiency suffers\cite{Bussi2015,Valsson2016,Pietrucci2017,Bussi2018}.
The exploration of the system is still enhanced, and the calculation does eventually converge\cite{Dama2014_convergence}, but this can require a substantial computational effort because, even after depositing the bias, some slow modes are present and the system retains a multiscale nature, with two separate timescales.
More in detail, in this case the shape of the free energy surface in the basins is easily reconstructed, while their relative height is much harder to converge.

Based on this observation we propose a multiscale approach that separates the calculation in two steps, and aims at improving efficiency in a suboptimal CV scenario.
First we reconstruct separately the free energy profile of each basin, using MetaD.
In a second step we use the properties of VES to optimize their free energy difference, and finally reconstruct the global FES.
The purpose of our approach is to take into consideration the multiscale nature of a suboptimally enhanced simulation to speed up convergence.

First we shortly present to the unfamiliar reader MetaD and VES, and we introduce a simple illustrative model that is helpful in visualizing the problem, then we explain in detail the proposed method.
Finally we successfully apply the new method to some representative systems.

\section{Metadynamics and variationally enhanced sampling}\label{S:recap}
In this section we wish to recall briefly the main features of the MetaD and VES methods, both used in our approach.
For a more in-depth review see e.g.~Ref.~\citenum{Valsson2016}.
Both methods enhance sampling by building on the fly a bias potential that is a function of a few collective variables. 
This bias discourages the system from remaining trapped in a metastable basin and forces it to explore other regions.
It also provides an estimate of the free energy, which is related to the probability distribution:
\begin{equation}
    \frac{e^{-\beta F(\mathbf{s})}}{Z}=P(\mathbf{s})
    \propto \int d\mathbf{R}\, e^{-\beta U(\mathbf{R})}\, \delta[\mathbf{s}-\mathbf{s}(\mathbf{R})]\, ,
\end{equation}
$Z=\int d\mathbf{s} e^{-\beta F(\mathbf{s})}$ being the normalization partition function, $U(\mathbf{R})$ the potential energy, and $\beta=1/(k_BT)$.

Metadynamics builds its bias as a sum of Gaussian contributions deposited periodically.
In particular we will use its well-tempered variant (WTMetaD)\cite{Barducci2008}, in which the height of the deposited Gaussian decreases exponentially as the bias is deposited:
\begin{equation}
    V_n(\mathbf{s})=\sum^n_{k=1}G(\mathbf{s},\mathbf{s}_k) 
    e^{ -\frac{\beta}{\gamma-1} V_{k-1}(\mathbf{s}_k)}\, ,
\end{equation}
where $G(\mathbf{s},\mathbf{s}_k)=We^{-\|\mathbf{s}-\mathbf{s}_k\|^2}$ is a Gaussian centered in $\mathbf{s}_k$, the CVs value at time $t_k$.
Independently from the choice of the Gaussian kernel parameters, this bias converges in the asymptotic limit to:
\begin{equation}
    V(\mathbf{s})=-(1-1/\gamma)F(\mathbf{s})+c\, ,
\end{equation}
where $F(\mathbf{s})$ is the FES and $c=c(t)$ does not depend on $\mathbf{s}$.
Once at convergence, the biased ensemble distribution  $P_V(\mathbf{s})\propto e^{-\beta [F(\mathbf{s})+V(\mathbf{s})]}$ is a smoother version of the unbiased one, i.e.~$P_V(\mathbf{s})=[P(\mathbf{s})]^{1/\gamma}$.
This broadening is controlled by the bias factor $\gamma$, which can go from $1$ (unbiased case) up to infinity (non-well-tempered MetaD).

In the VES method, on the other hand, the bias is obtained as the result of the minimization of a convex functional:
\begin{align}\label{E:omega}
  \Omega [V]=\frac{1}{\beta}
  \log \frac{\int d\mathbf{s}\, e^{-\beta [F(\mathbf{s})+V(\mathbf{s})]}}{\int d\mathbf{s}\ e^{-\beta F(\mathbf{s})}}
  +\int d\mathbf{s}\ p(\mathbf{s}) V(\mathbf{s})\, ,
\end{align}
where $p(\mathbf{s})$ is an arbitrary probability distribution called target distribution.
The VES functional is related to the relative entropy, or Kullback-Leibler divergence\cite{Invernizzi2017}.
The minimum condition is reached when $P_V(\mathbf{s})=p(\mathbf{s})$, a relation that can also be written as:
\begin{equation}\label{E:min_cond}
  F(\mathbf{s})=-V(\mathbf{s})-\frac{1}{\beta}\log p(\mathbf{s})\, .
\end{equation}
It is possible to choose as target distribution the well-tempered one, $p(\mathbf{s})=[P(\mathbf{s})]^{1/\gamma}$.
In this case the target has to be self-consistently adjusted as shown in Ref.~\citenum{Valsson2015}.
At convergence one obtains the same bias potential as WTMetaD.

In order to carry out the minimization of $\Omega[V]$, the bias is usually expanded over an orthogonal basis set $V(\mathbf{s})=\sum_k \alpha_k f_k(\mathbf{s})$, or parametrized according to some physically motivated FES model\cite{McCarty2016,Piaggi2016,Invernizzi2017}.
In this way the functional becomes a function of the variational parameters, and can be minimized using a stochastic gradient descent algorithm as described in Ref.~\citenum{Valsson2014}.

\section{An illustrative model}\label{S:model}
\begin{figure}
  \centering
  \includegraphics[width=0.5\columnwidth]{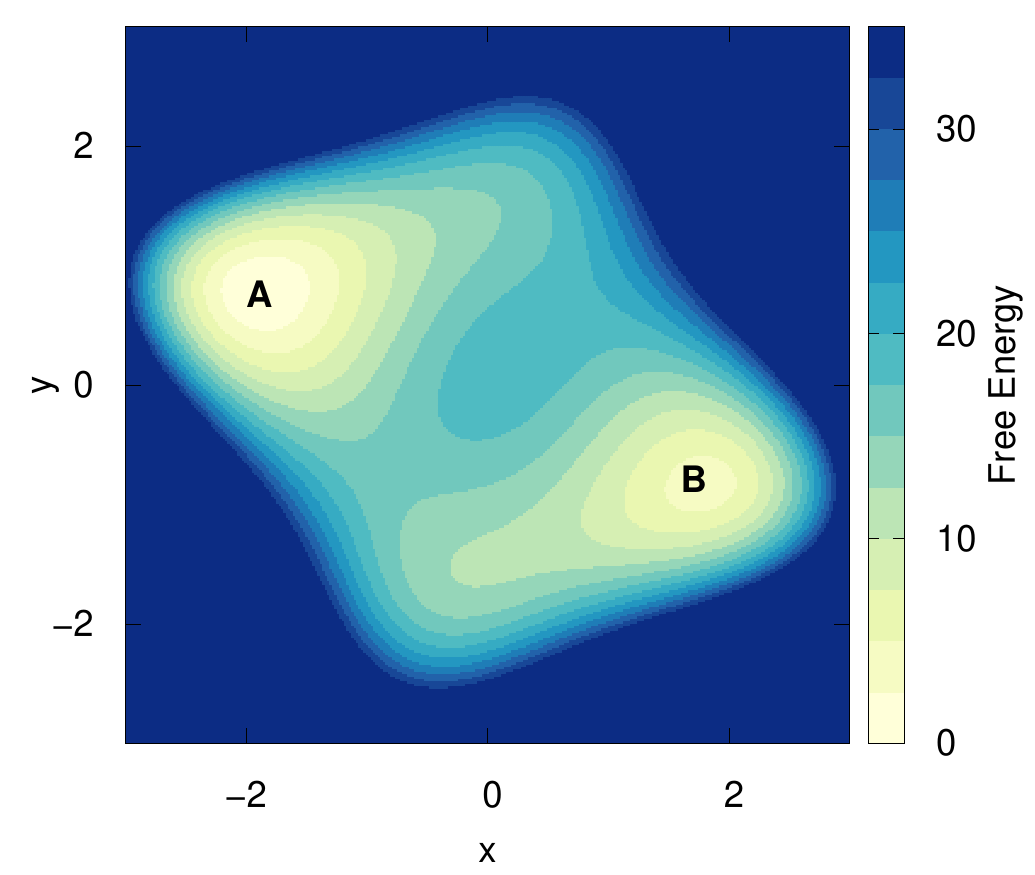}
  \caption{The 2D free energy surface of the considered model, with the two basins $A$ and $B$.}
  \label{F:model}
\end{figure}

\begin{figure}
  \centering
  \includegraphics[width=0.5\columnwidth]{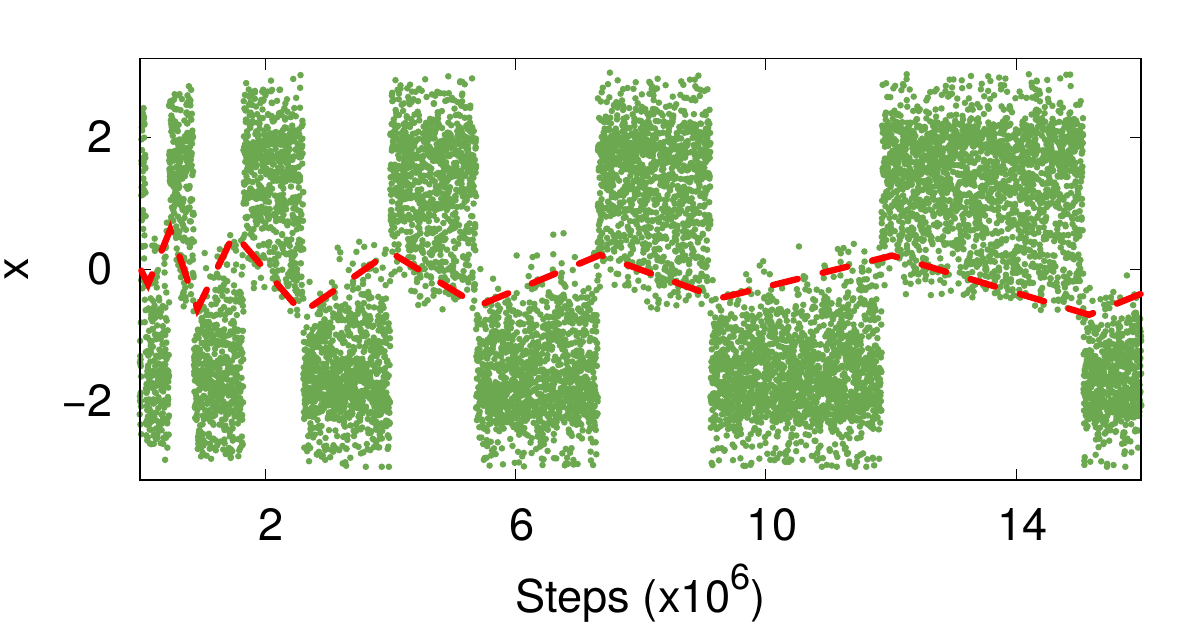}
  \caption{Typical trajectory of a suboptimal CV, obtained via a WTMetaD run ($\gamma=10$). 
    Hysteresis can clearly be seen (highlighted by the red dashed line), and it is also visible how WTMetaD slowly reduces it, by depositing less bias as the simulation proceeds.
    While it is not possible to draw an horizontal line that separates the two basins, one can easily determine the basin of belonging for each point by inspecting the CV time evolution.}
  \label{F:colvar}
\end{figure}

In order to better acquaint the reader with the problem described in the introduction, we illustrate the phenomenology associated with the use of suboptimal CVs in a simple model.
The model consists of a single particle moving with a Langevin dynamics on a 2D potential energy surface (see Fig.~\ref{F:model}) with two metastable basins, $A$ and $B$, such that transitions between them are rare.
This system has only two degrees of freedom, namely the $x$ and $y$ positions.
If we take $x$ as CV and perform a well-tempered MetaD run (details in the SI), we obtain the result in Fig.~\ref{F:colvar}.
The fact that we are missing one degree of freedom can be seen from the hysteresis apparent in the CV time evolution.
In order to go from one basin to the other, one must wait for a rare $y$ fluctuation, resulting in a delay that cannot be enhanced by adding a bias on $x$. 
Nevertheless, if we put more bias in one basin we open some higher pathways and transitions are observed.
However, they do not generally follow the lowest free energy path.
This extra bias will also prevent the system from coming back to the same basin until it is properly compensated as the MetaD simulation progresses, and this gives rise to hysteresis.

Well-tempering reduces the amount of bias deposited in such a way that the hysteresis will eventually disappear, and in the asymptotic limit the bias will converge\cite{Dama2014_convergence}.

If we monitor the time dependence of the bias $V(x)$, we can see how the shape of the two basins is learned after just the first few transitions, and remains pretty much unchanged throughout the simulation.
After this quick initial phase, the vast majority of the computational effort is used to pin down the free energy difference $\Delta  F$ between the two basins.

\section{Method}\label{S:method}
The observations made earlier suggest a two-step strategy.
First one obtains the shape of the free energy surface in the different metastable basins, then optimizes the free energy difference between them.
We will refer to this two-step procedure as VES$\Delta F$.

\subsection{Free energy surface model}
In a rare event scenario, such as the one discussed in Sec.~\ref{S:model}, there are by definition distinct metastable basins and it is thus possible to identify their separate contribution to the global FES.
Formally this can be done by considering the conditional probability $P(\mathbf{s}|A)$, i.e.~the probability that the CVs have value $\mathbf{s}$ when the system is in basin $A$.
\begin{equation}
    P(\mathbf{s}|A)\propto \int_A d\mathbf{R}\, e^{-\beta U(\mathbf{R})}\, \delta[\mathbf{s}-\mathbf{s}(\mathbf{R})]\, .
\end{equation}
From $P(\mathbf{s}|A)$ one then can write an associated free energy:
\begin{equation}
    \frac{e^{-\beta F_A(\mathbf{s})}}{Z_A}=P(\mathbf{s}|A)\, ,
\end{equation}
where $Z_A$ is a normalization constant.
With an identical procedure one can define $F_B(\mathbf{s}|B)$.

Notice that in general these probabilities are not mutually exclusive and, especially in the case of suboptimal CVs, there can be an overlapping region between them (see Fig.~\ref{F:colvar}).

The global probability distribution is then obtained simply by combining these conditional probabilities:
\begin{equation}\label{E:probability}
    P(\mathbf{s}) = P(\mathbf{s}|A) P_A + P(\mathbf{s}|B) P_B\, ,
\end{equation}
and the global free energy (modulo a constant) can thus be written as:
\begin{equation}\label{E:fes}
    F(\mathbf{s})=-\frac{1}{\beta}\log\left[\frac{e^{-\beta F_A(\mathbf{s})}}{Z_A}+\frac{e^{-\beta F_B(\mathbf{s})}}{Z_B}\, e^{-\beta \Delta F}\right]\, ,
\end{equation}
where the free energy difference between the basins $\Delta F$ is defined by:
\begin{equation}\label{E:deltaF}
    e^{-\beta \Delta F}=\frac{P_B}{P_A}=\frac{\int_B d\mathbf{R}\, e^{-\beta U(\mathbf{R})}}{\int_A  d\mathbf{R}\, e^{-\beta U(\mathbf{R})}}\, .
\end{equation}

One could also write the global FES as a function of $\Delta F_h$, defined as the difference in height between the two free energy minima.
This can be a fairly good and practical approximation to $\Delta F$, especially if the two basins have similar shapes.
More precisely, well within the typical statistical uncertainty (see SI), the two quantities differ by a constant that depends only on $F_A(\mathbf{s})$ and $F_B(\mathbf{s})$.
If the local free energies are shifted so that $\min[F_A(\mathbf{s})]=\min[F_B(\mathbf{s})]=0$, one can write:
\begin{equation}\label{E:delta_delta}
    \Delta F_h=\Delta F-\frac{1}{\beta}\log\frac{Z_A}{Z_B}\ ,
\end{equation}
and
\begin{equation}\label{E:fes_h}
    F(\mathbf{s})=-\frac{1}{\beta}\log\left[e^{-\beta F_A(\mathbf{s})}+e^{-\beta F_B(\mathbf{s})}e^{-\beta \Delta F_h}\right]\, ,
\end{equation}
with $\min[F(\mathbf{s})]=0$. 

\subsection{Step one: local basins}\label{SS:step1}
There are different ways of calculating the local free energies $F_A(\mathbf{s})$ and $F_B(\mathbf{s})$; here we suggest that a good estimate can be obtained using MetaD.
We do this by accumulating the bias up to the point in which the system escapes the basin from which the simulation was started.
As can be seen e.g.~in Fig.~\ref{F:colvar}, the first recrossing usually happens extremely fast, even if the CVs are suboptimal.
To increase the accuracy, WTMetaD can be used instead of plain MetaD, and also multiple runs can be combined, starting from different initial conditions. 

With this initial estimate it is important to be able to draw the two local FES up to the transition region, because this will ensure that they combine smoothly in Eq.~\eqref{E:fes}.
Such information is always obtained in the case of suboptimal CVs, thanks to the typical hysteresis.

In our approach we start $N$ independent simulations from basin $A$ and stop each simulation as soon as a transition occurs.
The bias $V_i(\mathbf{s})$ deposited by replica $i$ is then averaged with the others to obtain $V_A(\mathbf{s})=1/N\sum_iV_i(\mathbf{s})$ which is used to estimate the local free energy as  $F_A(\mathbf{s})=-(1-1/\gamma)^{-1}V_A(\mathbf{s})$, following the usual MetaD rules.
The same is done for basin $B$.

A very simple way to automatically stop the simulation at the first recrossing is to stop it when the CV reaches a threshold value set well beyond the transition region (e.g.~at the center of the other basin), and then cut a small segment at the end of the trajectory, to make sure that only configurations from the proper basin are kept.
Another option for detecting a recrossing is to use a descriptor different from the biased CV. 
The only requirement for such a descriptor is that it can partition the phase space in two regions, one for each basins, thus, contrary to a CV, it can be a discontinuous function of the atomistic coordinates, or even a function of past configurations, such as an exponentially decaying time average, which can greatly improve basins separation.

In case one has already performed an exploratory MetaD simulation with multiple transitions between the basins, another possibility for obtaining $F_A(\mathbf{s})$ and $F_B(\mathbf{s})$ is to separate the CVs trajectory and build a different reweighted free energy for each basin.

It should be noted that at this point an extremely precise determination of the free energy basins is not needed, since a more accurate description can be obtained later with a reweighting procedure of the longer convergence run.
In our experience, accuracy of the order of $k_BT$ in $F_A(\mathbf{s})$ and $F_B(\mathbf{s})$ does not slow down the overall convergence.

\subsection{Step two: free energy difference}\label{SS:step2}
Once the local basins are obtained we still need to know their relative free energy difference $\Delta F$ in order to build the global FES.
We use VES to estimate it.

As target distribution for VES we use the well-tempered one as in Ref.~\citenum{Valsson2015}, because it remains controllably close to the physical distribution, while at the same time enhancing the transition rate.
We can then write the bias potential expansion as a function of a single parameter, $\Delta F$, by combining Eq.~\eqref{E:fes} and \eqref{E:min_cond}:
\begin{equation}\label{E:bias}
  V(\mathbf{s})=(1-1/\gamma) 
  \frac{1}{\beta} \log \left[\frac{e^{-\beta F_A(\mathbf{s})}}{Z_A} + \frac{e^{-\beta F_B(\mathbf{s})}}{Z_B}\, e^{-\beta \Delta F}\right]\, .
\end{equation}
We can then minimize $\Omega$ as a function of $\Delta F$, while updating the estimate of the target distribution in a self-consistent way.

Given the suboptimal nature of the CVs, it was somehow natural to consider the introduction of a damping factor in the optimization algorithm commonly used in VES\cite{Bach2013}.
For this we took inspiration from the AdaGrad\cite{AdaGrad} stochastic gradient descent algorithm, generally used for training neural networks in a sparse reward scenario.
Further details can be found in the Supporting Information.

It is relevant to note that during the minimization procedure no extra bias is added, thus the system remains in the region of CVs space explored during step one and does not spend time unnecessarily in high free energy configurations.

\subsection{Reweighting procedure}\label{SS:reweighting}
The reweighting procedure is a key part of the VES$\Delta F$ method, because (as we will see) it converges faster than the VES optimization itself, and provides a more accurate estimate of $\Delta F$.
For reweighting we employ the scheme of Ref.~\citenum{Tiwary2015} and \citenum{Yang2018}, where the unbiased Boltzmann distribution $P(\mathbf{s})$ is obtained by sampling the biased one $P_V(\mathbf{s})$:
\begin{equation}
    P(\mathbf{s})= \frac{e^{-\beta F(\mathbf{s})}}{Z}=
    \frac{Z_V}{Z}\, e^{+\beta V(\mathbf{s})}\, \frac{e^{-\beta [F(\mathbf{s})+V(\mathbf{s})]}}{Z_V}
    =e^{-\beta[V(\mathbf{s})-c(t)]}P_V(\mathbf{s})\, ,
\end{equation}
where $Z_V=\int d\mathbf{s}\, e^{-\beta [F(\mathbf{s})+V(\mathbf{s})]}$ and the ratio $Z_V/Z=e^{-\beta c(t)}$ is estimated by numerical integration on a grid in the CVs space, using $F(\mathbf{s})=-(1-1/\gamma)^{-1}V(\mathbf{s})$.
This reweighting procedure can be used only in the adiabatic limit, when the applied bias is quasi-stationary.

When dealing with an optimal CV, a good biasing strategy is to quickly update the estimate of the underlying FES and correct accordingly the applied bias.
Since the CV incorporates all the relevant slow modes of the system, this strategy ensures a faster transition rate between the metastable basins and a faster convergence.
However, if the CV is suboptimal, as soon as we are close enough to a good FES estimate, the bottleneck for making a transition becomes the unbiased slow mode.
In such scenario, the choice of rapidly changing the applied bias is no more convenient, as it will only lead to hysteresis, not to faster convergence.
This is one of the reasons why well-tempered metadynamics outperforms plain metadynamics in many applications, and why the ``first fill, then converge'' strategy of transition-tempered metadynamics\cite{Dama2014_ttmetad} can be so effective.

In VES$\Delta F$ we push this strategy even further, and we design the bias optimization in such a way that it quickly reaches a value close to the optimal one, and from then one we only make gentle adjustments.
This allows us to maximize the time spent in the adiabatic limit, thus improving the reweighting efficiency.
As can be seen e.g.~in Fig.~\ref{F:model_deltaF}, the result is that the estimate coming from the reweighting can reach convergence faster than the direct $\Delta F$ optimization.

The other aspect that makes reweighting a convenient tool is accuracy.
We do not spend much time in step one, thus the shape of the local basins may not be very accurate. 
In principle this can lead to a systematic error in the estimate of $\Delta F$, since it depends not only on the relative height of the basins $\Delta F_h$ but also on their shapes, Eq.~\eqref{E:delta_delta}.
Thus a safer way of estimating $\Delta F$ is to obtain a reweighted $F(\mathbf{s})$ and to explicitly integrate in the CVs space:
\begin{equation}
    \Delta F=-\frac{1}{\beta} \log
    \left[\frac{\int_B d\mathbf{s}\, e^{-\beta F(\mathbf{s})}}{\int_A d\mathbf{s}\, e^{-\beta F(\mathbf{s})}}\right]\, .
\end{equation}
Such an estimate is based on the much longer step-two run, and is a very good approximation provided that the CVs are able to distinguish the two basins, which is usually the case even for suboptimal CVs.
This is why in the code implementation we prefer to use the FES model of Eq.~\eqref{E:fes_h}, and optimize $\Omega$ with respect to $\Delta F_h$, and then use reweighting to estimate $\Delta F$.

\section{Results}\label{S:results}
The VES$\Delta F$ method has been implemented in PLUMED\cite{plumed}, an enhanced sampling plug-in which we used in combination with Gromacs\cite{gromacs} and LAMMPS\cite{lammps} molecular dynamics packages.
Our code is openly available online\bibnote{Currently in the development version of PLUMED (master branch on GitHub), under the VES module, with the name \texttt{VES\_DELTA\_F}. The provided implementation can deal with multidimensional CVs and more than two FES basins.}.

\begin{figure}
  \centering
  \includegraphics[width=0.5\columnwidth]{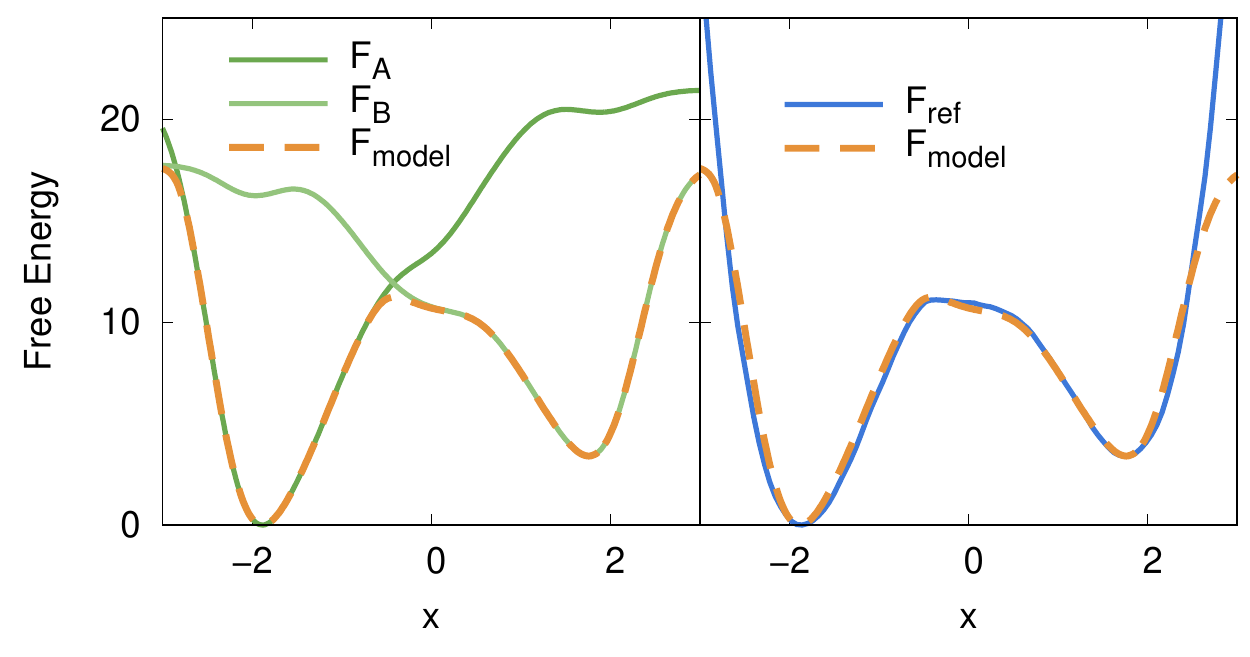}
  \caption{On the left the local free energies $F_A(x)$ and $F_B(x)$ are compared to the model one $F_{model}(x)$ obtained by combining them following Eq.~\eqref{E:fes_h}. 
    On the right the same model free energy is compared to the reference one $F_{ref}(x)$, that can be obtained via the reweighting procedure.
    The parameter $\Delta F_h$ is unknown, and is calculated during the VES optimization.}
  \label{F:free_energies}
\end{figure}
\begin{figure}
  \centering
  \includegraphics[width=0.5\columnwidth]{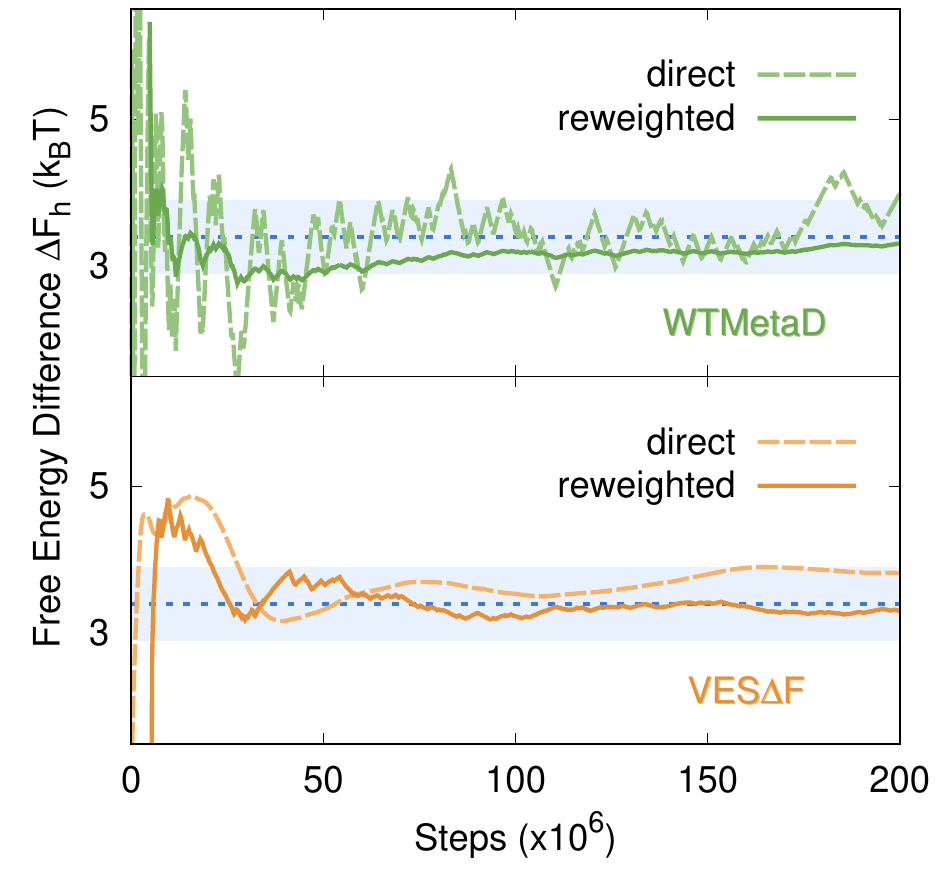}
  \caption{Estimate for the illustrative model of Sec.~\ref{S:model} of $\Delta F_h$ along $x$, obtained directly from the acting bias and from the reweighting procedure (Sec.~\ref{SS:reweighting}), in the case of WTMetaD and VES$\Delta F$. 
    In both methods a bias factor $\gamma = 10$ is used.
    The reference blue stripe is $1\, k_BT$ thick.}
  \label{F:model_deltaF}
\end{figure}
First we apply our approach to the case of the simple model of Sec.~\ref{S:model}.
In order to build the local free energy basins we follow step one (Sec.~\ref{SS:step1}) and run five independent replicas using WTMetaD with $\gamma=5$ enhancing $x$ fluctuations.
We obtain the $F_A(x)$ and $F_B(x)$ shown in Fig.~\ref{F:free_energies}, with a computational effort that is negligible in comparison to the one needed to converge the global FES.

We use these local basins to perform step two and optimize the free energy difference through VES ($\gamma=10$, $\mu=0.05$).
Fig.~\ref{F:model_deltaF} shows an example of such optimization, compared to a WTMetaD run.
We also show the estimate of $\Delta F_h$ obtained with the reweighting procedure.

Both WTMetaD and VES$\Delta F$ reach relatively quickly a rough estimate of the $\Delta F_h$ between the basins, but while the bias used by the former keeps oscillating, the latter behaves more smoothly.
We believe this is ultimately the reason why the reweighting converges faster in the case of VES$\Delta F$.

\subsection{Alanine dipeptide}\label{SS:alanine}
The Alanine dipeptide molecule in vacuum is often used as benchmark for enhanced sampling methods.
At $300$ K it presents two main metastable basins, separated by a kinetic bottleneck.
The most stable one, $A$, includes two different conformations, C5 and C7eq, that are separated only by a small barrier.
Basin $B$ instead, hosts only one metastable conformation, known as C7ax.

\begin{figure}
  \centering
  \includegraphics[width=0.5\columnwidth]{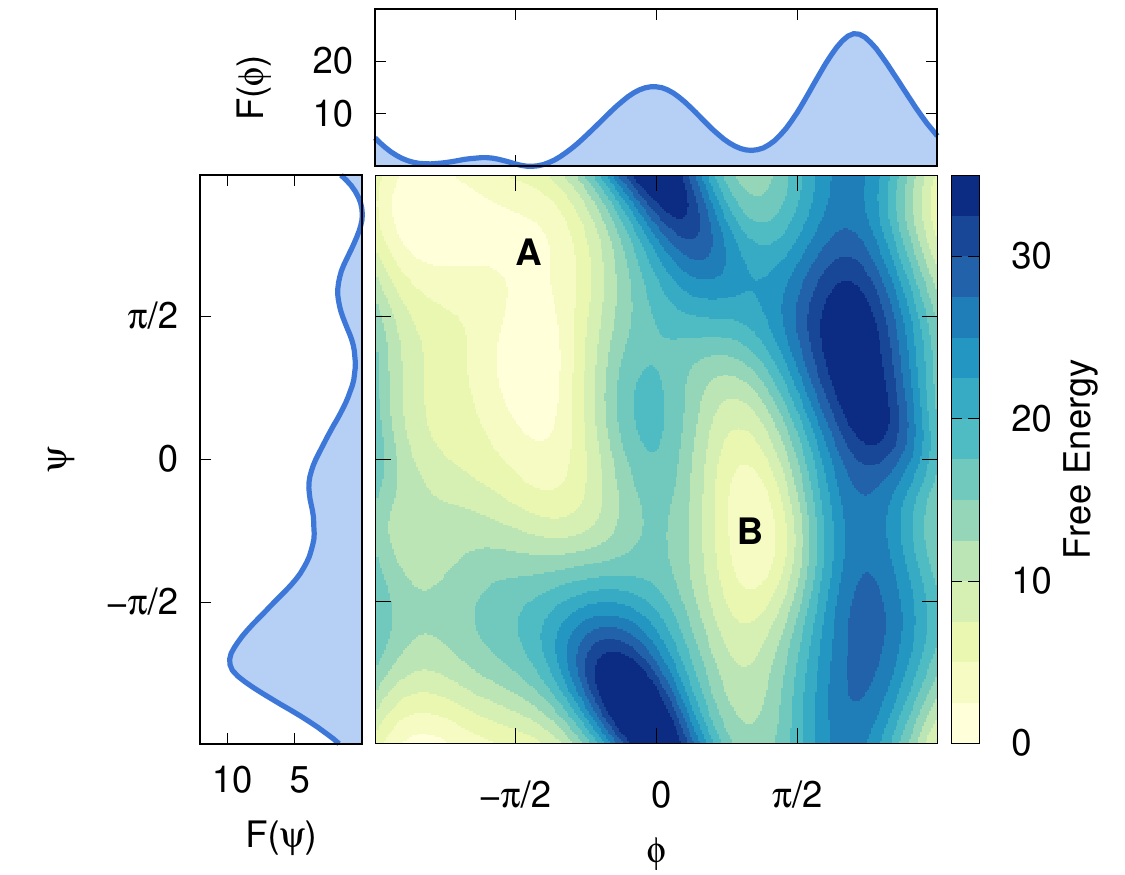}
  \caption{The free energy surface of alanine dipeptide in vacuum, as a function of the $\phi$ and $\psi$ angles.
    On the sides is the free energy projected onto each single angle.}
  \label{F:ala_FES}
\end{figure}
The typical CVs used for this system are the two Ramachandran angles $\phi$ and $\psi$, Fig.~\ref{F:ala_FES}.
The first one, $\phi$, is an almost optimal CV, which allows for very fast convergence ($\sim 10$~ns) when used in MetaD.
Instead $\psi$ is not only suboptimal, but also a typical example of a bad CV.
When biasing $\psi$, transitions between the basins are enhanced, but the system remains multiscale, with a significant timescale gap between intra-basin and inter-basins fluctuations.
In this scenario MetaD presents a strong hysteresis and its convergence is rather slow.
For the purpose of showing the strength of our method, we will pretend here that we do not know of the existence of the $\phi$ angle, and we make the unfortunate choice of biasing only $\psi$.

It is important to recall that while the free energy surface as a function of $\psi$ is very different from the one relative to $\phi$ (and thus the minima relative height $\Delta F_h$ is different), the free energy difference $\Delta F$ between the two basins does not depend on the CV used, and is the same independently of the CV representation of the FES, along $\psi$, $\phi$ or both (see SI).

In order to reconstruct the local free energies we run 10 independent WTMetaD simulations for each basin, terminating them as soon as they make the first transition.
We use a bias factor $\gamma=10$, and other simulation details can be found in the Supporting Information.
\begin{figure}
  \centering
  \includegraphics[width=0.5\columnwidth]{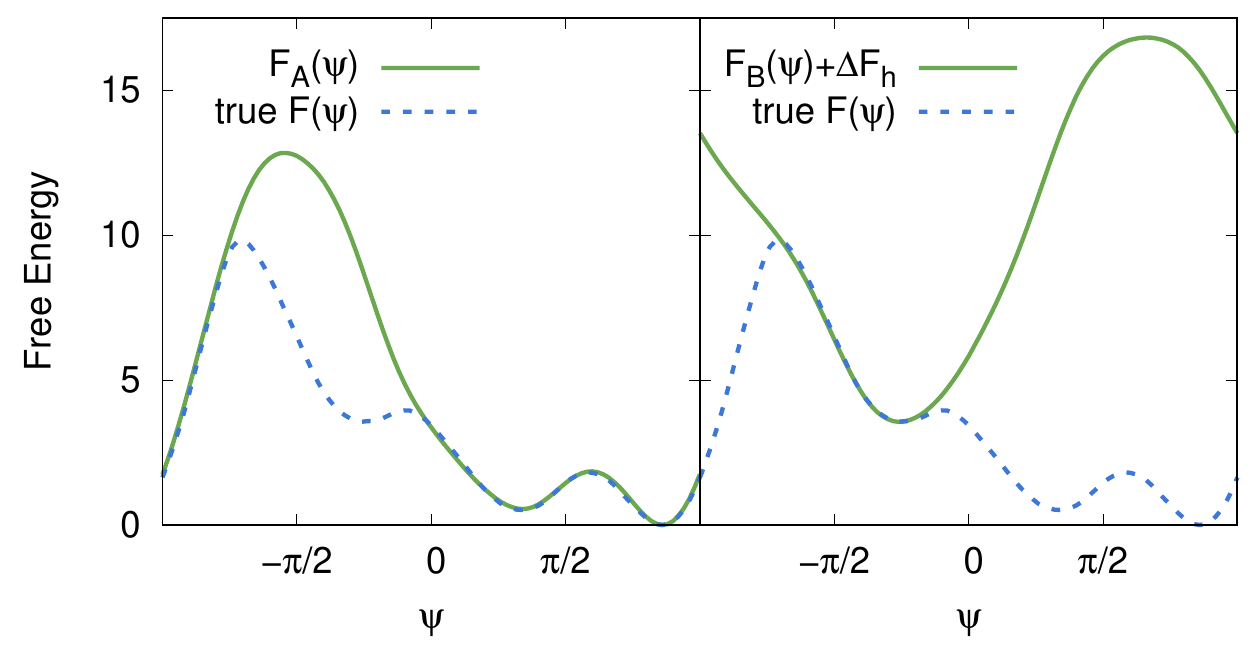}
  \caption{The local free energies obtained for alanine dipeptide.
    As reference a fully converged $F(\psi)$ is used.
    The quantity $\Delta F_h$ is unknown at this stage, but is used here for displaying purposes.}
  \label{F:alanine_local_FES}
\end{figure}
On average it takes $0.55$~ns to exit the most stable basin $A$, and $0.16$~ns to escape basin $B$.
The combined total simulation time employed for determining $F_A(\psi)$ and $F_B(\psi)$ is about $7$~ns.
This provides a very good estimate, as can be seen in Fig.~\ref{F:alanine_local_FES}. 
For the $\Delta F_h$ optimization we use a target distribution with a well-tempered bias factor $\gamma=10$, and a minimization step $\mu=0.05$.
As a reference, we perform runs of WTMetaD and transition tempered metadynamics\cite{Dama2014_ttmetad} (TTMetaD), with $\gamma=10$ (see SI).
We also perform simulations with different numbers of multiple walkers\cite{walkers}.

\begin{figure}
  \centering
  \includegraphics[width=0.5\columnwidth]{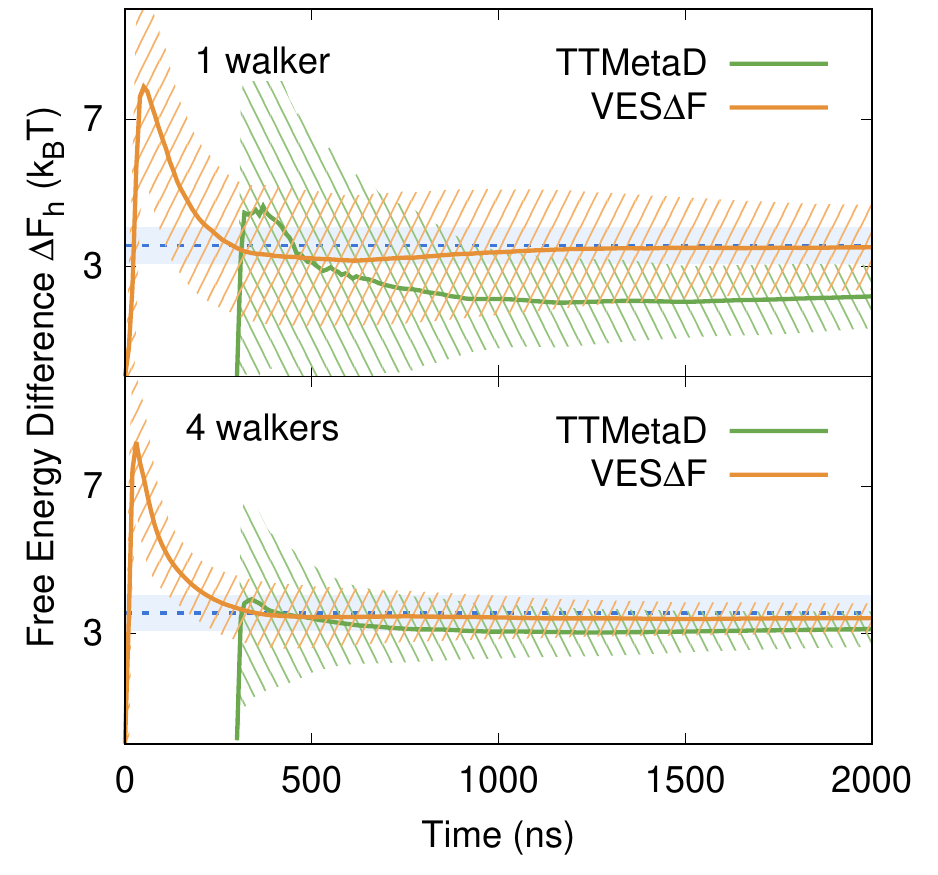}
  \caption{Comparison between alanine $\Delta F_h$ convergence using TTMetaD and VES$\Delta F$, with 1 and 4 multiple walkers ($\gamma=10)$.
  The $\Delta F_h$ is obtained through the reweighting procedure described in Sec.~\ref{SS:reweighting}.
  In the case of MetaD we must exclude from the reweighting an initial transient, in which the system is out of equilibrium and the estimate of $c(t)$ is unreliable.
  We run 100 independent replicas starting from basin $A$ with different initial conditions.
  Only the average and the standard deviation is shown.
  The reference blue stripe is $1\, k_BT$ thick.}
  \label{F:alanine_stats}
\end{figure}

In order to have an estimate of the uncertainty on the calculated $\Delta F_h$, we run for each of the three methods multiple completely independent runs, always starting from basin $A$, but with different initial conditions.
We then look at the average and the standard deviation of these replicas, as a function of time.
Some of the results are shown in Fig.~\ref{F:alanine_stats}, and more can be found in the Supporting Information.

On average TTMetaD performs better than plain WTMetaD, while VES$\Delta F$ is a great improvement over either method.
In particular the VES$\Delta F$ reweighting is more accurate even in the case of a single walker, where error bars are large and the optimization is not yet at convergence.
MetaD methods show a systematic shift which is not compatible with the estimates coming from the bias and disappears when more statistics are collected (the same happens in Fig.~\ref{F:sodium_deltaF}).

Increasing the number of walkers makes sampling of the neglected slow degrees of freedom more efficient and improves convergence.
Our method, VES$\Delta F$, scales particularly well with the number of multiple walkers.

Finally we would like to stress the fact that we are not suggesting to purposely pick bad CVs, because even if our method brings better results when compared to MetaD, a simulation based on $\phi$ instead of $\psi$ would still converge with a computational effort smaller by orders of magnitude.

\subsection{Sodium}\label{SS:sodium} 
As a last example we study the liquid-solid transition of sodium at $350$ K.
It is a first order phase transition, and the stable solid structure is a body centered cubic (bcc) lattice.

Contrary to the previous case, for this example we use a CV that is one of the best available to describe the system.
We consider a recently developed CV\cite{Niu2018} based on the peaks of the Debye structure factor (see SI).
Such a CV is able to clearly identify the liquid and the solid phase, and can drive transitions between the two.
Nevertheless it is not an optimal CV, and has some hysteresis, as can be seen e.g.~in Fig.~3 of Ref.~\citenum{Niu2018}.

Although suboptimal, this CV is able in our system to drive the transitions between liquid and solid avoiding defected states, which might introduce secondary metastable minima in the FES, and affect the estimate of the liquid-solid  $\Delta F$.

\begin{figure}
  \centering
  \includegraphics[width=0.5\columnwidth]{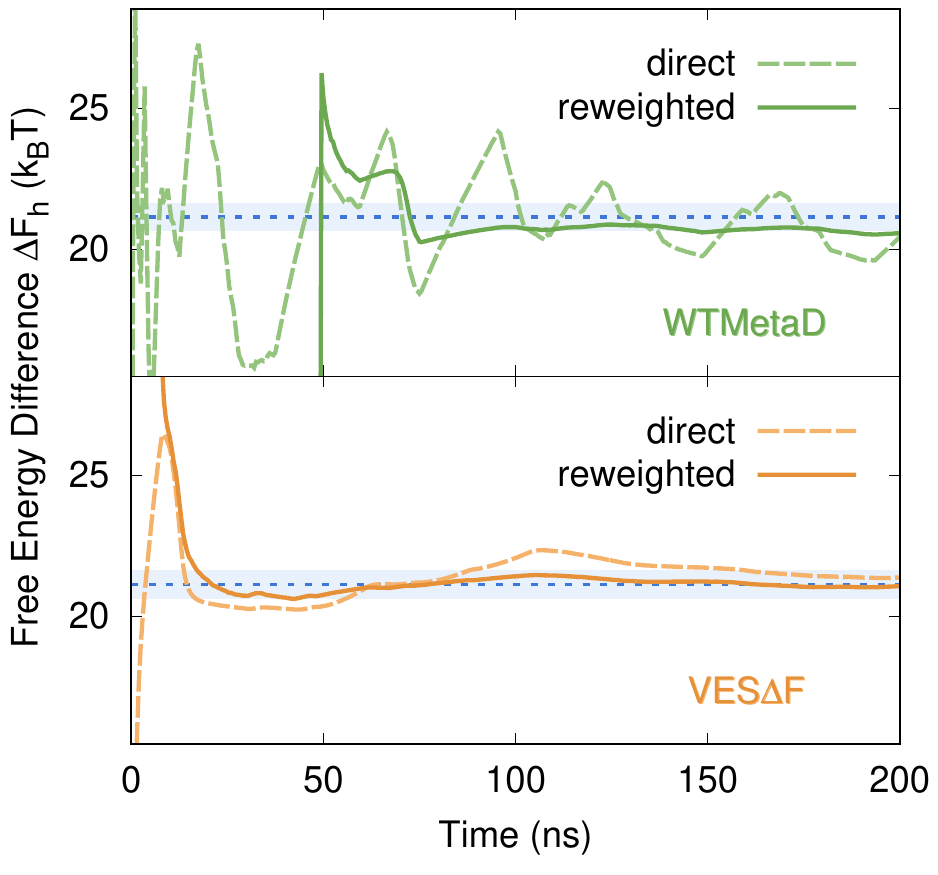}
  \caption{Convergence of $\Delta F_h$ between liquid and solid for sodium at $350$ K, obtained directly from the acting bias and from the reweighting procedure (Sec.~\ref{SS:reweighting}), using WTMetaD and VES$\Delta F$. 
    In both methods a bias factor $\gamma = 20$ and 4 multiple walkers were used.
    The reference blue stripe is $1\, k_BT$ thick.}
  \label{F:sodium_deltaF}
\end{figure}

Similarly to what was done in the previous case, we first build the local free energy basins by running 5 short WTMetaD runs ($\gamma =20$), then use them to optimize the free energy difference.
The target distribution is a well-tempered one with $\gamma =20$ and the minimization step is $\mu=1$.
In Fig.~\ref{F:sodium_deltaF} we show the convergence of VES$\Delta F$ in comparison with WTMetaD.
Both simulations run with 4 parallel multiple walkers; see SI for more details.

Also in this less extreme case of a suboptimal CV, VES$\Delta F$ outperforms the standard approaches.

\section{Conclusions}\label{S:conclusions}
In this paper we propose a new method, based on a combination of MetaD and VES, to calculate the free energy difference between two metastable states.
In order to perform such a calculation MetaD and VES are in principle very appropriate, especially when used with an optimal CV that is able to accelerate all the slow modes of the system.
In such a scenario the computational effort needed for convergence is close to that needed for exploring the FES landscape.

However, in real life applications suboptimal CVs are used, and this makes it useful to separate the free energy reconstruction into two steps.
First we use MetaD to explore the basins, and then a bespoke version of VES to converge the free energy difference.
In doing so, our method focuses on approaching a quasi-stationary bias, which improves the efficiency of the reweighting procedure and thus the convergence speed.

In some cases one does not know in advance if the available CV is suboptimal or not, and which are the relevant basins of the system under study.
A typical usage scenario would then be to run a first rough non well-tempered MetaD simulation, aimed at discovering the relevant free energy basins and the presence (or absence) of hysteresis.
If this preliminary run reveals the presence of a big number of secondary metastable minima, as it can happen e.g.~when dealing with defects in a phase transitions, then VES$\Delta F$ would not be the best choice.

We would like to notice that in a suboptimal scenario the VES optimization is crucial, and the proper $\Delta F$ cannot be retrieved via some reweighting technique, such as weighted histogram analysis (WHAM\cite{Kumar1992}), applied to the MetaD simulations in the basins.
This is because the overlap observed in the CV space is not actually an overlap in the phase space.
This issue would not be solved by simply adding new biased windows (in an umbrella sampling spirit), since the hysteresis would still be present\cite{Zhu2012}.

The present method, VES$\Delta F$, is to some extent similar to that proposed in Ref.~\citenum{McCarty2016}, but its motivations are different, as are the FES model used and the strategy to obtain the local basins.

Previous authors have proposed variants to the MetaD algorithm to improve its convergence rate\cite{Barducci2008,Dama2014_ttmetad,Branduardi2012}, but none of them connects directly to the source of the problem, namely the residual multiscale behaviour of suboptimal CVs.
We think this is a simple but important consideration, and our approach is meant to fully take into account the multiscale nature of suboptimal CVs enhanced sampling.

Another important feature is that it is very transparent, having only $\Delta F$ as a parameter to optimize, that is the very physical quantity one is interested in.
It also focuses only on the relevant part of the CV space, avoiding exploring new regions during the convergence phase.
This simplicity can be very helpful, especially when dealing with complex systems.

Furthermore, VES$\Delta F$ scales very well in the case of multiple walkers, making good use of parallel simulations.
For the sake of simplicity we presented here only the case of one dimensional CVs and two minima systems, but the implementation of VES$\Delta F$ provided in the publicly available  PLUMED code can already deal with multidimensional CVs and more than two basins.
Our work points to a useful strategy that can be applied also in other circumstances and with other methods.

\begin{acknowledgement}

This research was supported by the NCCR MARVEL, funded by the Swiss National Science Foundation, and European Union Grant No.~ERC-2014-AdG-670227/VARMET. 
Calculations were carried out on Euler cluster at ETH Z\"urich and on M\"onch cluster at the Swiss National Supercomputing Center.

The authors thank Emanuele Grifoni for useful discussions, GiovanniMaria Piccini for precious help with the artworks, and Omar Valsson for carefully reading the manuscript.

\end{acknowledgement}

\begin{suppinfo}

The Supporting Information is available free of charge on the ACS Publications website at https://pubs.acs.org/doi/suppl/???????.
It contains:
\begin{itemize}
    \item Sampling with different CVs
    \item Notes on Eq.~\eqref{E:fes_h} and \eqref{E:delta_delta}
    \item Optimization algorithm: AdaGrad damping and hyperparameters choice
    \item Illustrative model: computational details
    \item Alanine dipeptide: computational details and more results
    \item Sodium: computational details
\end{itemize}

\end{suppinfo}

\bibliography{ves_deltaF_Biblio}

\begin{mcitethebibliography}{37}
\bibitem[Not()]{Note-1}
Currently in the development version of PLUMED (master branch on GitHub), under
  the VES module, with the name \texttt{VES\_DELTA\_F}. The provided
  implementation can deal with multidimensional CVs and more than two FES
  basins.\relax
\mciteBstWouldAddEndPunctfalse
\mciteSetBstMidEndSepPunct{\mcitedefaultmidpunct}
{}{\mcitedefaultseppunct}\relax
\EndOfBibitem
\end{mcitethebibliography}

\end{document}